\documentclass[aps,twocolumn,showkeys,groupedaddress,amsmath,nofootinbib,amssymb,showpacs,floatfix]{revtex4-1}
\usepackage{graphicx}  % needed for figures
\usepackage{dcolumn}   % needed for some tables
\usepackage{color}
\usepackage{psfrag}
\usepackage{bbm,bm}
\usepackage{epsfig}
\usepackage{hyperref}
\allowdisplaybreaks
\begin{document}
\def\be{\begin{equation}}
\def\ee{\end{equation}}
\def\bea{\begin{eqnarray}}
\def\eea{\end{eqnarray}}
\def\ba{\begin{array}}
\def\ea{\end{array}}
\def\ben{\begin{enumerate}}
\def\een{\end{enumerate}}
\def\nab{\bigtriangledown}
\def\tpi{\tilde\Phi}
\def\nnu{\nonumber}
\newcommand{\eqn}[1]{(\ref{#1})}
\def\bw{\begin{widetext}}
\def\ew{\end{widetext}}
\newcommand{\half}{{\frac{1}{2}}}
\newcommand{\vs}[1]{\vspace{#1 mm}}
\newcommand{\dsl}{\pa \kern-0.5em /} 
\def\a{\alpha}
\def\b{\beta}
\def\g{\gamma}\def\G{\Gamma}
\def\d{\delta}\def\D{\Delta}
\def\ep{\epsilon}
\def\et{\eta}
\def\z{\zeta}
\def\t{\theta}\def\T{\Theta}
\def\l{\lambda}\def\L{\Lambda}
\def\m{\mu}
\def\f{\phi}\def\F{\Phi}
\def\n{\nu}
\def\p{\psi}\def\P{\Psi}
\def\r{\rho}
\def\s{\sigma}\def\S{\Sigma}
\def\ta{\tau}
\def\x{\chi}
\def\o{\omega}\def\O{\Omega}
\def\k{u}
\def\pa {\partial}
\def\ov{\over}
\def\nn{\nonumber\\}
\def\ud{\underline}
\def\ct{\textcolor{red}{\it cite }}
\def\qq{\text{$Q$-$\bar{Q}$ }}

%\nofiles

%\preprint{APS/123-QED}

\title{\large{\bf Confinement and Pseudoscalar Glueball Spectrum in the $2+1$D\\ QCD-Like Theory from the Non-Susy D$2$ Brane}}
\author{Adrita Chakraborty}
\email{adimanta09@iitkgp.ac.in}
\affiliation{
   Indian Institute of Technology Kharagpur,\\
   Kharagpur 721302, India}
\author{Kuntal Nayek}
\email{kuntal.nayek@iitkgp.ac.in}
\affiliation{
   Indian Institute of Technology Kharagpur,\\
   Kharagpur 721302, India}
\date{\today}

\begin{abstract}
Here we study two important properties of $2+1$ dimensional QCD -- confinement and pseudoscalar glueball spectrum -- with holographic approach. We consider the low energy decoupled geometry of the isotropic non-susy D$2$ brane. We find the corresponding gauge theory is similar to the $2+1$-dim Yang-Mills theory with the running coupling $\lambda^2$. At the extremal limit (i.e. BPS limit), this gauge theory reduces to the super-YM theory. From the Nambu-Goto action of a test string, the potential of a \qq pair located on the boundary is calculated. At large \qq separation it gives the tension $\sigma$ of the QCD flux-tube. The parametric dependence of $\sigma$ is shown pictorially. It is found that $\sigma$ is a monotonically increasing function of the effective coupling $\lambda^2$. In comparison, $\sqrt{\sigma}/g_\text{YM}^2N_c$ is found to match accurately with the previous results. In the next part, we consider fluctuation of the axion field in the aforementioned gravity background. From the linearized field equation of the fluctuation we calculate the mass spectrum of $0^{-+}$ numerically using the WKB approximation. The pseudoscalar mass is found to be related to the string tension approximately as $M_{0^{-+}}/\sqrt{\sigma}\approx 3(n+2)$ for first three energy states, $n=0,\,1,\,2$.   
\end{abstract}

%\pacs{11.25.-w, 11.25.Tq, 11.25.Uv}

%Use showkeys class option if keyword
\maketitle

\tableofcontents

\section{Introduction}

According to our current understandings through various scattering experiments in RHIC, LHC etc, most of the physical phenomena are mainly explained with the $3+1$ dimensional non-Abelian gauge theory -- QCD. Since last five decades, the perturbative picture of this theory is almost well-established with the proper theoretical models as well as the experimental evidences. There are some theoretical approaches like MIT bag model, Skyrme model, chiral perturbation theory, heavy baryon perturbation theory for the strong coupling regime of QCD. However, this regime of the $3+1$ dimensional QCD has not been explored enough compared to the perturbative regime.  The lattice QCD has some non-trivial approach in this regime. Although the accuracy of the lattice calculation is restricted with various technical limitations and it is also unable to study the real-time dynamics of the theory. The lattice calculation with finite lattice-spacing is approximated to the continuum limit to map the real QCD. But this process is not always easy and accurate. Since last two decades, the string theory has become relevant to study the strongly coupled gauge theory through the holographic tool, called AdS/CFT correspondence \cite{Maldacena:1997re,Aharony:1999ti}. In this correspondence, the non-perturbative regime of the supersymmetric Yang-Mills theory in $3+1$ dimension has been studied almost completely. But in reality our known QCD is not supersymmetric. So, still it is interesting to study the non-perturbative QCD in $3+1$ dimensions.

Addressing this issue, the non-pertubative QCD in lower dimension, e.g., $2+1$ dimensional Yang-Mills theory has been studied \cite{Karabali:1995ps,Teper:1998te} due to the following reasons. Unlike the four dimensional theory, it is comparatively simple to handle mathematically. However unlike the four dimensional theory, this theory having the dimension-full coupling $g_\text{YM}^2$ is not classically scale-invariant. In spite of this difference, the most important reason to study this lower dimensional theory is that there are many conceptual similarities with the $3+1$ dimensional Yang-Mills theory. The theory has ultraviolet freedom like four dimensional theory. The dimensionless effective coupling of the theory $g_\text{YM}^2\ell_s$ goes to zero at the ultra high energy scale, i.e., at the small length scale $\ell_s\to 0$. Also the infra-red slavery, confinement and mass spectrum have same characteristic as four dimensional QCD. At low energy scale the coupling becomes too large and both the theories are driven by self-coupling of the gluon. However the $2+1$ dimensional Coulomb potential has weak logarithmic confinement $V_C(r)\sim g_\text{YM}^2\log(r)$, the linear confinement can be found to have the well-known form at the non-perturbative regime, i.e., $V(r)\sim\sigma r$. In $2+1$ dimensional pure Yang-Mills theory, the string tension of the QCD flux-tube is theoretically \cite{Nair:2002yg} derived as,
\be\label{tensionref}
\sigma=g_\text{YM}^4\frac{N_c^2-1}{8\pi}
\ee
for large $N_c$. This direct proportionality of the binding energy with the spatial scale indicates the confined state of the bounded \qq pair. This confinement indicates the self-coupling dominated non-perturbative regime. In this regime, due to the strong self-coupling, the free gluons form the bound states, called glueball. Like the four dimensional theory, here the masses of the light glueball states are also found to be proportional to $\sqrt{\sigma}$. Along with these theoretical approaches, the three dimensional YM theory has been studied in lattice simulation \cite{Teper:1998te,Teper:1993gm,Philipsen:1996af}. However for the three dimensional theory, the available data are not enough to get the complete information about the mass spectrum of the pseudoscalar glueball. 

The AdS/CFT correspondence \cite{Maldacena:1997re,Aharony:1999ti} states the duality between the supergravity theory in the $4+1$ dimensional anti-de Sitter space-time and the $\mathcal{N}=4$ Super-Yang-Mills theory in $3+1$ dimensional flat space-time. According to this correspondence, a stack of $N_c$ BPS D$3$ branes in the gravity theory is used to study the dual pure super-YM theory. In the gravity theory, the string coupling $g_s$ is so small that the theory is perturbative. On the other hand, in the gauge theory, however the Yang-Mills coupling $g_\text{YM}^2$ is small ($g_\text{YM}^2\sim g_s\to0$), the number of the gauge fields $N_c$ is large, $N_c\to\infty$. The effective coupling of the gauge theory, $g_\text{YM}^2N_c$, is finite and large enough to make the theory non-perturbative \cite{Maldacena:1997re}. So this duality is useful to study the non-perturbative quantum field theory with the large effective coupling through the holographic study of the perturbative gravity theory. Thus it is also called strong/weak coupling duality. However the AdS/CFT duality is restricted to the $\mathcal{N}=4$ super Yang-Mills theory on the AdS boundary, the same idea of the general gauge/gravity duality can be extended to the non-AdS gravity theory to study the strongly coupled Yang-Mills theory without the conformal symmetry \cite{Witten:1998zw}. 
In these holographic models, the effective gauge coupling is large which does not allow the theory to be asymptotically free \cite{Polchinski:2001tt}. Despite the absence of asymptotic freedom, many other non-perturbative behaviours of QCD like confinement, chiral symmetry breaking, scattering cross-section, glueballs, thermal phase transition, etc., can be studied from holography \cite{Witten:1998zw,Constable:1999ch,Ooguri:1998hq,Csaki:1998qr,Babington:2003vm,Csaki:2006ji,Kim:2007qk,Polchinski:2002jw}. The four dimensional QCD has been also studied in the light-front holographic approach \cite{Brodsky:2010ur} and in lattice calculations \cite{Morningstar:1999rf,Teper:1997tq,Miller:2006hr,Lucini:2001ej}. The general concept of gauge/gravity duality is also used in the lower dimensional branes where the dual theories are the lower dimensional QCD-like theories. In holographic approach, like four dimensional gauge theory, the asymptotic freedom is not allowed in QCD$3$. But it allows to study other properties of QCD. The QCD$3$ has been already studied in holographic method \cite{Aharony:1999ti,Csaki:1998qr,Hong:2010sb} by compactifying the D$3$ brane in one of its spatial directions to get the $3+1$ dimensional non-AdS gravity theory. 

Here, in this article we intend to study three dimensional QCD-like theory from the non-BPS D$2$ brane of the type-II supergravity. The confinement property and the mass spectrum of the pseudoscalar glueball are studied in this work. In ten dimensional type-II supergravity theory there is a set of non-supersymmetric solutions \cite{Lu:2004ms} which does not follow the BPS condition. As the supersymmetry is broken the solutions have more parameters than BPS solutions. In dual non-conformal gauge theory, these parameters are related to the coupling and the UV fixed point which make the theory QCD-like. Here we consider a stack of $N_c$ number of the non-susy, isotropic D$2$ branes where the background dilaton field $\phi$ varies with the length scale of the bulk theory. According to gauge/gravity duality, it corresponds to $2+1$ dimensional Yang-Mills theory with running effective coupling which is analogous to $2+1$ dimensional QCD. The low energy decoupling limit is considered to find the near brane geometry of the non-susy D$2$ brane. Then we configure a probe string whose end points (representing the \qq pair on $t-x^1$ plane) are separated by a large distance $\Delta x$ along one of the spatial directions of the brane $x^1$. Following the holographic dictionary, the thermal expectation value of the time-like Wilson loop $W^F(\mathcal{C})$ is computed from the minimal world-sheet area swept out by the test string on the boundary surface of the gravity theory which coincides with the loop $\mathcal{C}$ \cite{Aharony:1999ti,Constable:1999ch,Liu:2006he}. Hence we use the following relation \eqref{wilson} to find the \qq potential energy $V(\Delta x)$ from the minimal Nambu-Goto action $S(\mathcal{C})$.
\be\label{wilson}
W^F(\mathcal{C})=\text{Exp}[iS(\mathcal{C})]=\text{Exp}[iV(\mathcal{C})\mathcal{T}]
\ee 
where $\mathcal{T}$ is the length of the temporal direction of the loop $\mathcal{C}$. For large separation, the \qq potential is found to be linearly proportional to the separation; $V(\Delta x)\propto\Delta x$ which ensures the confinement nature of the gauge theory. In the holographic approach, the mass spectrum can be evaluated from the fluctuations in the gravity theory. According to the properties of the fluctuated fields we get different types of glueballs, e.g. scalar, vector and tensor fields respectively correspond to the spin-$0$, spin-$1$ and higher spin glueballs. The naked singularity of the non-susy D$2$ does not allow us to find the scalar glueball with the particular method used herein. However this singularity should not be an issue if we approach it with some other method. So here we study the pseudoscalar glueball only. To study this glueball, we consider the fluctuation of the axion field in the same decoupled background. As the axion is the minimally coupled scalar, the fluctuation does not perturb the background metric. From the potential energy term in the Schr\"odinger-like field equation of the fluctuation, we compute the mass spectrum of the pseudoscalar glueball. The mass spectrum is evaluated numerically using the WKB approximation. The restrictions in the WKB method bound us in the lower states only. In the non-perturbative regime, since the effective coupling is very large the theory is dominated by the self-coupling of the gluon fields. So in this regime we can compare our results with the Yang-Mills theory without the flavour quark. The pseudoscalar glueball has spin zero, odd parity and even charge parity; $0^{-+}$. So far the estimated ground state mass of the pseudoscalar glueball $0^{-+}$ is $2590$MeV \cite{Morningstar:1999rf} and $M_{-+}^*/M_{-+}=1.46$ \cite{Aharony:1999ti} in $3+1$ dimension. In four dimension, the glueball mass is found to increase with the gauge coupling. The same feature is expected in lower dimension too. 

This article is arranged as follow. In the section II, we discuss the non-susy D$2$ brane and its decoupling limit. After that in the following section, the confinement is studied from the \qq potential using a test string. Then the mass spectrum is calculated and the result is compared with the known data in the section IV. Finally in the last section we conclude the study. \\

\section{Gravity Background}

In this section we will discuss the non-susy D$2$ brane solution and its decoupled geometry at low energy limit.

\subsection{Non-Susy D$2$ Brane}
Besides the supersymmetric BPS brane, there is a similar sector of the non-supersymmetric solutions of the type-II supergravity, which are called, in short, non-susy D$p$ brane \cite{Lu:2004ms}. Unlike the BPS D$p$ brane, these are non-extremal solutions, i.e., the ADM mass and charge are not equal rather follow the gravitational censorship. Due to this non-equality, the non-susy branes are characterized by more parameters. In the BPS limit, which is indeed the extremal limit, the number of free parameters reduces and the non-susy branes merge into BPS branes of same dimensions. At low energy limit, the non-susy branes decouple from the bulk theory in similar manner as BPS branes do \cite{Nayek:2015tta,Roy:2017mje}. But in these cases, the decoupled geometries are not AdS or conformal to AdS. It means the decoupled geometry is not symmetric under the conformal transformations. According to the gauge/gravity duality, this non-AdS gravity theory corresponds to the non-conformal gauge theory defined in the worldvolume of the brane. Again the dilaton field associated to this non-susy brane is found to be non-trivial and depends on the length scale of the gravity theory. So in holographic dictionary, the dual gauge theories in case of these non-susy D$p$ branes are similar to the non-supersymetric Yang-Mills theory with the running coupling.

The non-susy D$2$ brane solution can be derived by putting $p=2$ and re-arranging the harmonic function $F(\rho)$ in $(2.11)-(2.12)$ of \cite{Nayek:2015tta}.
\bea\label{geometry}
ds^2 &=& F(\rho)^{-\half}G(\rho)^{\frac{\b}{4}+\frac{\d}{4}}\left(- dt^2 
+ \sum_{i=1}^2 (dx^i)^2\right)\nn
& & +F(\rho)^\half G(\rho)^{\frac{1}{5}-\frac{\b}{4}+\frac{\d}{4}}\left(\frac{d\rho^2}{G(\rho)} + \rho^2 d\Omega_6^2\right)\nn           
e^{2\phi} &=& e^{2\phi_0} F(\rho)^\half G(\rho)^{\d - \frac{\b}{4}},\quad F_{[6]} = Q \text{Vol}(\Omega_6)
\eea
where the harmonic functions are
\bea
F(\rho) & = & G(\rho)^\g\cosh^2\theta-\sinh^2\theta\nn
G(\rho) & = & 1+\frac{\rho_2^5}{\rho^5}
\eea
Here the background is given in the String frame. The $(2+1)$ dimensional worldvolume is defined with the coordinates $(t,x^1,x^2)$ whereas the seven dimensional transverse space is defined by the spherical coordinates $(\rho,\Omega_6)$. $\rho$ is the radial coordinate perpendicular to the worldvolume, i.e., the brane is located at $\rho=0$. Therefore the geometry has a singularity at $\rho=0$. The dilaton field $\phi$ is the function of radial coordinate. The boundary expectation value of dilaton, $\phi_0$, is related to the string coupling $g_s=e^{\phi_0}$. As the effective coupling of the theory depends on the dilaton field, it also varies with the length scale $\rho$ of the theory or the energy scale ($\sim 1/\rho$). $\rho_2$ is a constant point on the length scale, also known as the mass parameter. $\theta$ is a dimensionless constant, known as the charge parameter, related to the total RR charge $Q$ of the brane. Along with these three, there are three more dimensionless parameters -- $\d,\,\b$ and $\g$. These six parameters are mutually related via the following three relations derived from \cite{Nayek:2015tta}.
\bea
&& \b = \g +\frac{\d}{4}\nn
&& \g = \sqrt{\frac{12}{5}-\frac{15}{16}\d^2}\nn
&& Q = \frac{5}{2}\g\rho_2^5\sinh 2\theta\label{parameter}
\eea
Therefore, out of the six we have only three independent free parameters -- $\rho_2,\,\theta$ and $\d$. Now according to the definition, the harmonic function $G(\rho)$ is always greater than $1$. Validity of the metric in the range $0<\rho<\infty$ refers $F(\rho)$ to be positive, which suggests that $\g$ must be positive and real. So the value of the parameter $\d$ is bounded in the range $-\frac{8}{5}\leq\d\leq\frac{8}{5}$. 

In the BPS limit \cite{Lu:2004ms}, which is eventually the extremal limit, $\rho_2\to 0$ and $\theta\to\infty$, but $Q$ remains finite and reduces to the RR charge of the BPS D$2$ brane; $Q=5R_2^5$, where $R_2$ is the mass parameter of the BPS D$2$ brane. Therefore in this limit $G(\rho)\sim 1$ and $F(\rho)$ reduces to the harmonic function of BPS D$2$ brane, i.e., $F(\rho)\to 1+\frac{R_2^5}{\rho^5}$. Thus the background \eqref{geometry} reduces to the BPS D$2$ brane.

\subsection{Decoupled Geometry}

Now, to study the gauge/gravity duality in this non-susy background, at the low energy limit, we must have two completely decoupled theories -- theory with gravity and theory without gravity. The theory without gravity, i.e., the gauge theory lives on the brane. This decoupling occurs when the gravitational excitations living in the bulk decouple from the theory on the brane. This has been already confirmed in previous studies from the graviton scattering cross-section \cite{Nayek:2015tta}. The graviton scattering in this background at low energy limit shows that the near brane regime decouples from the ten dimensional bulk. Here we are interested in the near brane decoupled geometry of non-susy D$2$ brane. As the fundamental string length $\ell_s\to 0$ at the low energy limit, we consider the length scale of the bulk accordingly to get a finite scale in near brane regime. Now following the decoupled geometry of the non-susy D$3$ brane \cite{Nayek:2016hsi}, we scale the radial coordinate and parameters of \eqref{geometry} in the following way.
\be\label{dcscale}
\rho=\a'u,\,\rho_2=\a'u_2,\,\sinh^2\theta=\frac{L}{\g u_2^5\a'^2}
\ee
where $\a'=\ell_s^2$ has mass dimension $-2$. Since $\rho$ has mass dimension $-1$, $u$ has mass dimension $+1$. $u$ is taken as the radial coordinate of the decoupled geometry which is identified as the energy scale of the theory. $u_2$ is a constant parameter on the energy scale $u$, which is the cause of the non-conformal structure of the theory. The other quantity $L$ is defined as $L=3\pi^2 g_\text{YM}^2N_c$, where $g_\text{YM}^2$ is the Yang-Mills coupling in $2+1$ dimensional Yang-Mills theory and $N_c$ is the color charges. In gravity theory, Yang-Mills coupling is related to the string coupling $g_\text{YM}^2=2g_s\a'^{-\half}$ and $N_c$ is the number of the D$2$ branes. As $g_s$ is dimensionless, the Yang-Mills coupling has mass dimension $1$. As $\a'\to 0$, the string coupling $g_s$ has to be very small too so that the dilaton field does not diverge near the singularity $\rho=0$. This also makes the gauge coupling small enough. To satisfy the gauge/gravity duality, as $g_\text{YM}^2$ is small, $N_c$ must be very large. Thus using the scaling \eqref{dcscale}, we can write the decoupled geometry of the background \eqref{geometry} as follows.
\bea
ds^2 & = & \a'G_{\mu\nu}dX^\mu dX^\nu\nn 
& = & \a'\left[\sqrt{\frac{\g u_2^5}{L}}F(u)^{-\half}G(u)^{\frac{\d}{4}+\frac{\b}{4}}\left(-dt^2+\sum_{i=1}^2(dx^i)^2\right)\right.\nn
&& \left.+\sqrt{\frac{L}{\g u_2^5}}F(u)^\half G(u)^{\frac{1}{5}+\frac{\d}{4}-\frac{\b}{4}}\left(\frac{du^2}{G(u)}+u^2d\Omega_6^2\right)\right]\nn
e^{2\phi} & = & \frac{g_\text{YM}^4}{4}\sqrt{\frac{L}{\g u_2^5}}F(u)^\half G(u)^{\d-\frac{\b}{4}}\nn
F_{[6]} & = & 5L\a'^3\text{vol}(\Omega_6)\label{dcgeometry}
\eea
where the harmonic functions get the new form under the decoupling limit \eqref{dcscale} as,
\be 
F(u)=G(u)^\g-1,\quad G(u)=1+\frac{u_2^5}{u^5}
\ee 
In these decoupled forms of the harmonic functions, $\g$ can not be zero. So the allowed range of $\d$ is now modified to $-\frac{8}{5}<\d<\frac{8}{5}$. This decoupled geometry is non-AdS, there is no conformal symmetry. Now as we have considered $L$ to be very large, the curvature of the six dimensional transverse sphere is small enough at the finite energy scale which validates the application of the gauge/gravity duality. In \eqref{dcgeometry}, the corresponding $(2+1)$ dimensional gauge theory on the boundary is expected to be non-conformal theory due to the presence of $u_2$ in the gravity theory. As we take $u_2\to 0$, the metric becomes AdS with a non-trivial conformal factor same as we get it from BPS D$2$ brane at low energy limit. So $u_2$ is related to the fixed energy scale in the corresponding gauge theory. Now the effective gauge coupling $\lambda^2$ can be written from the dilaton field following the standard relation $e^\phi\sim\frac{\lambda^\frac{5}{2}}{N_c}$ \cite{Aharony:1999ti}.
\be\label{lambda} 
\lambda^2 = \frac{1}{(6\pi^2)^\frac{4}{5}\g^\frac{1}{5}}\left(\frac{L}{u_2}\right)F(u)^\frac{1}{5} G(u)^{\frac{2}{5}\d-\frac{\b}{10}}
\ee
which varies with the energy scale $u$ and the parameter $\d$. Thus in the corresponding gauge theory, we have the running coupling similar to QCD. So the theory on the brane is a $(2+1)$ dimensional QCD-like theory without flavour. Since the Yang-Mills coupling is directly proportional to $g_s$ and inversely proportional to the fundamental string length, at low energy limit in supergravity theory it remains finite and non-zero which does not allow the theory to be asymptotically free. 
\begin{figure}
    \centering
    \includegraphics[width=0.4\textwidth,height=4.5cm]{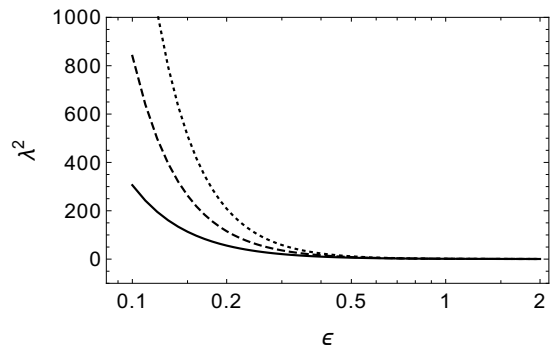}
    \caption{Dependence of $\lambda^2$ on the energy scale $u=\epsilon u_2$ for $\d=0.98\,\text{(solid line)},\,1.28\,\text{(dashed line)}$ and $1.58\,\text{(dotted line)}$ at $u_2=g_\text{YM}^2N_c$.}
    \label{lambda vs u}
\end{figure}
Like the QCD theory, in \eqref{lambda}, the effective gauge coupling $\lambda^2$ decays with the increasing energy scale $u=\epsilon u_2$ which is shown in the Figure \ref{lambda vs u}. Here we have plotted $\lambda^2$ vs $u$ for three different $\d$ values, $\d=0.98,\,1.28,\,1.58$. $\lambda$ is found to have large value at small $u$ and diverges at the singularity $u=0$. Near the singularity $\lambda^2\gg1$ which is the self-coupling dominated regime. In the asymptotic region $u\to\infty$, the coupling constant goes to zero mathematically. But in the holographic study the $\lambda<1$ regime is not allowed. At the fixed energy scale $u$, $\lambda^2$ is also found to increase monotonically with the increasing value of $\d$.

Now at this point we can understand the significance of the dimension-full parameter $u_2$ from the behaviour of the theory on the various energy scales $u$. At $u\gg u_2$, the geometry \eqref{dcgeometry} reduces to the decoupled geometry of the BPS D$2$ brane and the effective coupling becomes 
\be\label{bpslambda}
\lambda^2\approx \frac{(6\pi^2)^{\frac{1}{5}}}{2}\frac{g_\text{YM}^2N_c}{u}\sim \frac{g_\text{YM}^2N_c}{u}.
\ee
As $\frac{(6\pi^2)^{\frac{1}{5}}}{2}=1.1$ we can consider it as $1$. The corresponding gauge theory for the BPS D$2$ brane has been discussed in details in \cite{Aharony:1999ti}. In this case gauge theory is perturbative if
\be\label{limit1}
u\gg g_\text{YM}^2N_c. 
\ee
On the other hand, at the limit $u\ll u_2$, the effective coupling $\lambda^2$ becomes very large for the given range of $\d$. In this regime the effective string coupling $e^\phi>>1$ when 
\be\label{limit2}
u^{\frac{\g}{2}+\frac{15}{8}\d}\ll\frac{1}{\gamma^{1/5}}g_\text{YM}^2N_c^{\frac{1}{5}}u_2^{\frac{\g}{2}+\frac{15}{8}\d-1}. 
\ee
Then we need to uplift the non-susy D$2$ brane to the non-susy M$2$ brane solution defined in eleven dimensions.
Therefore the worldvolume theory of the non-susy D$2$ brane is holographically dual to the type IIA supergravity in the range $u^{\frac{\g}{2}+\frac{15}{8}\d}>\frac{1}{\gamma^{1/5}}g_\text{YM}^2N_c^{\frac{1}{5}}u_2^{\frac{\g}{2}+\frac{15}{8}\d-1}$ and $u<g_\text{YM}^2N_c$.
Now the $2+1$ dimensional Yang-Mills theory have only one dimension-full parameter which is the coupling $g_\text{YM}^2$. So in this picture, the bulk parameter $u_2$ is related to the coupling $g_\text{YM}^2$. Now combining $u>>u_2$ with \eqref{limit1}, $u_2$ can have a maximum value $g_\text{YM}^2N_c$. Again from the limits $u\ll u_2$ and \eqref{limit2}, $u_2$ can have a lower limit $\frac{1}{\gamma^{1/5}}g_\text{YM}^2N_c^{\frac{1}{5}}$. So according to holography, the type-IIA supergavity describes the non-perturbative gauge theory for the above range of $u$, only if
\be
\frac{1}{\gamma^{1/5}}g_\text{YM}^2N_c^{\frac{1}{5}}\leq u_2\leq g_\text{YM}^2N_c
\ee
Since from the above analysis, the maximum value of $u$ can be order of $g_\text{YM}^2N_c$, if we take $u_2\ll g_\text{YM}^2N_c$, such upper limit of $u$ results $u\gg u_2$ which leads the theory to the worldvolume theory of the BPS D$2$ brane. So for $u_2=\frac{1}{\gamma^{1/5}}g_\text{YM}^2N_c^\frac{1}{5}$ we get the world volume theory of BPS D$2$ brane for the whole range of $u$.
Here we take $u_2=g_\text{YM}^2N_c$. This choice ensures that we are entirely in the non-perturbative worldvolume theory of the non-susy D$2$ brane for the above range of $u$. In other words, we do our calculations in non-perturbative regime near the perturbative boundary.

\section{Confinement}

\begin{figure}
    \centering
    \includegraphics[width=0.4\textwidth,height=4.5cm]{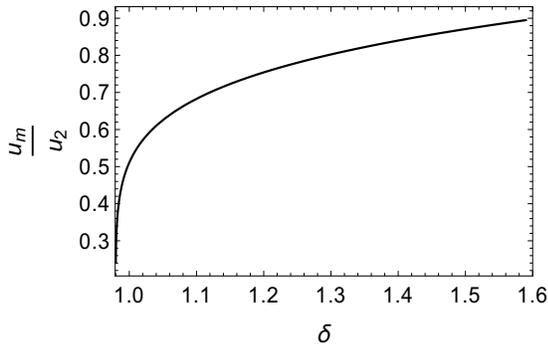}
    \caption{Dependence of the ratio $\frac{u_{m}}{u_{2}}$ with $\d$.}
    \label{u vs delta} 
    \end{figure}

\begin{figure}[t]
    \centering
    \includegraphics[width=0.4\textwidth,height=4.5cm]{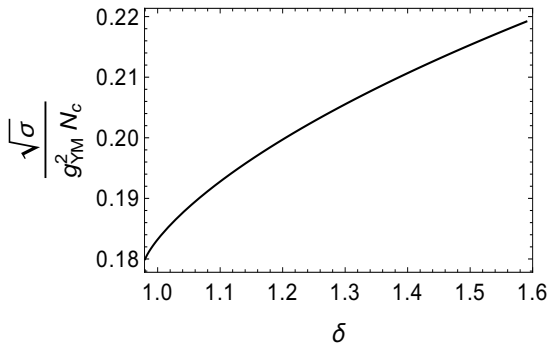}
    \caption{$\delta$-dependence of QCD string tension $\sigma$ at $u_2=g_\text{YM}^2N_c$. }
    \label{sigma vs delta}
\end{figure}

Now to study the confinement property we need to find the binding potential of a largely separated static \qq pair. To do that we consider a quark antiquark pair on the boundary of the gravity background and an open string connecting the pair is hanging towards the singularity $u=0$ in the bulk. The two dimensional worldsheet swept by this open string (test string) is compared with the Wilson loop in the flat boundary metric which gives the \qq potential using \eqref{wilson}. The area of the Wilson loop is given by the Nambu-Goto action, in the decoupled geometry, which can be written as follows.
\be\label{ngaction}
S=\frac{1}{2\pi}\int d^2\sigma\sqrt{-\text{Det}\left(g_{\a\b}\right)}
\ee
The action is calculated on two dimensional worldsheet governed by the coordinates $(\sigma^0,\sigma^1)$. The metric $g_{\a\b}$ is the pull-back of the bulk geometry on the worldsheet. This two dimensional pull-back metric is  $g_{\a\b}=G_{\mu\nu}\frac{\partial X^\mu}{\partial\sigma^\a}\frac{\partial X^\nu}{\partial\sigma^\b}$, where $X^\mu$ and $\sigma^\a$ are respectively bulk and worldsheet coordinates. Here we choose the test string with the following parametrization.
\be
\sigma^0\equiv t,\,\sigma^1\equiv x^1=x\,(\text{say}),\,u=u(x)
\ee 
where the other bulk coordinates remain localised at constant values. $x$ denotes the \qq separation which depends on the vertical length of the test string inside the bulk. With this parametrisation we derive the two dimensional worldsheet metric.
\bea
g_{\s^0\s^0} & = & G_{tt}\nn 
g_{\s^0\s^1} & = & g_{\s^1\s^0} = 0\nn
g_{\s^1\s^1} & = & G_{x^1x^1}+G_{uu}\left(\frac{du}{dx}\right)^2\nonumber
\eea 
Therefore, the action integral \eqref{ngaction} takes the following form.
\be
S=\frac{1}{2\pi}\int dt dx\,P(u)\left[1+M(u)^2\left(\frac{du}{dx}\right)^2\right]^\half 
\ee
where
\bea 
P(u) & = & \sqrt{-G_{tt}G_{x^1x^1}}=\sqrt{\frac{\g u_2^5}{L}}F(u)^{-\half}G(u)^{\frac{\d}{4}+\frac{\b}{4}}\label{pu}\\
M(u) & = & \sqrt{G_{uu}G^{x^1x^1}}=\sqrt{\frac{L}{\g u_2^5}}F(u)^\half G(u)^{-\frac{2}{5}-\frac{\b}{4}}\label{mu}
\eea 
The action depends on the slope of $u(x)$ -- the evolution of the string inside the bulk, $du/dx$, i.e., the rate of change of the distance $u$ of the string from the singularity with the \qq separation $x$. This slope vanishes at the stable configuration (stable position of the turning point) of the string. Now as the \qq separation increases the turning point moves toward the singularity, i.e., $u$ decreases. So the position of the turning point is minimum at the maximum separation. First we go to the stable point of the string by setting $du/dx = 0$, then we find the lowest position of that point taking the global minimum of $P(u)$. In this configuration, the maximum \qq separation is found to have the turning point at $u=u_m$,
\be\label{tpoint}
u_m=\frac{u_2}{\left(\Sigma^{1/\g}-1\right)^{1/5}}\quad\text{where,}\Sigma =\frac{5\d+4\g}{5\d-4\g}
\ee 
It is clear from the above turning point that it does not exist for the whole parametric range. Since $\g$ is always positive, it exists only if $\Sigma>1$ or $5\d+4\g>5\d-4\g$. Then the allowed parametric regime is now $\frac{2\sqrt{6}}{5}<\d<\frac{8}{5}$. Beyond this range, the test string goes into the singularity without any valid turning point. The variation of the ratio $\frac{u_{m}}{u_{2}}$ with the parameter $\delta$ in this specified parametric range is depicted in the Figure \ref{u vs delta}.
Now from the Wilson loop equation \eqref{wilson}, the binding energy or the potential is related to the area of the loop as $V=S/\mathcal{T}$, where $\mathcal{T}$ is the length of the temporal side of the loop $\mathcal{C}$, i.e., time runs from $0$ to $\mathcal{T}$. Thus the binding potential of the \qq pair at the maximum separation $\Delta x$ (i.e. at $u=u_m$) is 
\bea 
V & = & \frac{1}{2\pi}\int dx\, P(u_m)\nn 
%& = & \frac{u_2^\frac{5}{2}}{2\pi\sqrt{L}}\sqrt{\frac{5\d-4\g}{8}}\Sigma^{\frac{1}{4}+\frac{5\d}{16\g}}\Delta x\nn
%& = & \frac{u_2^\frac{5}{2}}{2\pi\sqrt{L}}\left[\frac{25}{64}\d^2-\frac{1}{4}\g^2\right]^\frac{1}{4}\Sigma^\frac{5\d}{16\g}\Delta x\\
& = & \frac{u_2^\frac{5}{2}}{2\pi\sqrt{L}}\left[\frac{5}{8}\d^2-\frac{3}{5}\right]^\frac{1}{4}\Sigma^\frac{5\d}{16\g}\Delta x\label{potential}
\eea
where $\int dx=\Delta x$ is the length of the flux-tube.
In \eqref{potential}, the \qq potential is linearly proportional to this spatial length. Coefficient of $\Delta x$ has dimension of string tension, which is identified as the QCD string tension $\sigma$. From \eqref{potential}, we can extract $\sigma$ as,
\be\label{tension}
%\sigma =\frac{u_2^\frac{5}{2}}{2\pi\sqrt{L}}\sqrt{\frac{5\d-4\g}{8}}\Sigma^{\frac{1}{4}+\frac{5\d}{16\g}}
\frac{\sigma}{g_\text{YM}^4N_c^2}=\frac{9\pi^3}{2}\left(\frac{u_2}{L}\right)^\frac{5}{2}\left(\frac{5}{8}\d^2-\frac{3}{5}\right)^\frac{1}{4}\Sigma^\frac{5\d}{16\g}
\ee 
Here in $(2+1)$ dimensional theory, the flux-tube tension depends on the fixed energy scale $u_2$, the dimension-full Yang-Mills coupling $g_\text{YM}^2N_c$ and on the background parameter $\d$. Again the effective gauge coupling $\lambda^2$ at $u=u_m$ also depends on these quantities.
\be  
\lambda^2(u_m)  =  \frac{1}{(6\pi^2)^\frac{4}{5}}\left(\frac{L}{u_2}\right)\left[\frac{5}{8}\d^2-\frac{3}{5}\right]^{-\frac{1}{10}}\Sigma^\frac{3\d}{8\g}
\ee
However the $\d$ dependence profile of $\lambda^2(u_m)$ is not similar as $\lambda^2(u)$. For a fixed $u$, $\lambda^2(u)$ is a monotonically increasing function of $\d$ and for a fixed $\d$, it is a decreasing function of $u$. We have also seen $u_m$ increases with the increasing value of $\d$. So $\lambda^2(u_m)$ varies with the combined effect of these two. As a result of this dependence $\lambda^2(u_m)$ is found to decrease with $\d$. Since $\lambda^2(u_m)$ is a dimensionless constant and $u_2$ has the same dimension as $L$, so in \eqref{tension}, $\sigma$ is dimensionally proportional to $g_\text{YM}^4N_c^2$, i.e., $\sqrt{\sigma}$ has dimension of mass. Now we replace $u_2$ in \eqref{tension} and re-write the expression in terms of the dimensionless quantity $\lambda(u_m)$. 
\be\label{tension1}
\sigma=\frac{(g_\text{YM}^2N_c)^2}{8\pi\lambda^5(u_m)}\Sigma^\frac{5\d}{4\g}=\frac{1}{2\pi\a'}\frac{(g_sN_c)^2}{\lambda^5(u_m)}\Sigma^\frac{5\d}{4\g}
\ee 
So the dependence of $\sigma$ as $\sqrt{\sigma}\propto g_\text{YM}^2N_c$ is similar to \eqref{tensionref} which validates the expression in \eqref{tension1}. Since $\a'\to 0$ because of decoupling limit and $g_s\to 0$ to get the perturbative gravity theory, the ratio $g_s/\a'$ is finite. Again as the gauge theory is strongly coupled, $\lambda$ is large, but the ratio $N_c^2/\lambda^5$ is finite due to large $N_c$. Therefore the QCD string tension has a finite value. $\delta$ dependence of $\sigma$ is given in Table \ref{tab:tension}. Here we have taken $u_2=g_\text{YM}^2N_c$ which makes the ratio $\frac{\sqrt{\sigma}}{g_\text{YM}^2N_c}$ independent of $g_\text{YM}^2N_c$. 
\begin{table}[h]
    \centering
    \caption{Values of $\sqrt{\sigma}$ calculated from \eqref{tension1} at $u_2=g_\text{YM}^2N_c$.}
    \begin{tabular}{|c|c|c|c|c|c|c|c|}
        \hline
         $\d$ & $0.98$ & $1.08$ & $1.18$ & $1.28$ & $1.38$ & $1.48$ & $1.58$ \\
         \hline
         $\frac{\sqrt{\sigma}}{g_\text{YM}^2N_c}$ & $0.1799$ & $0.1912$ & $0.1984$ & $0.2044$ & $0.2097$ & $0.2144$ & $0.2187$ \\ 
         \hline
    \end{tabular}
    \label{tab:tension}
\end{table}
Now the theoretical value of $\sqrt{\sigma}$ is $0.1995$\footnote{At $N_c\to\infty$, one can expand \eqref{tensionref} as $$\frac{\sqrt{\sigma}}{g_\text{YM}^2N_c}=\frac{1}{\sqrt{8\pi}}\left(1-\frac{1}{2N_c^2}+\frac{1}{8N_c^4}+\cdots\right)$$ So at $N_c=\infty$, $\sqrt{\sigma}/g_\text{YM}^2N_c=\frac{1}{\sqrt{8\pi}}=0.1995$.} for pure Yang-Mills theory. In lattice calculation \cite{Teper:1998te}, the value is $0.1975$. Both of these values are evaluated at $N_c=\infty$. Here for various $\d$, $\sqrt{\sigma}/g_\text{YM}^2N_c$ varies in the range $0.19\pm0.02$ for large $N_c$. So our calculation fairly matches with those previous results. It is also clear from the Table \ref{tab:tension} that the string tension is a monotonically increasing function of $\d$ which is depicted in Figure \ref{sigma vs delta}. As we have seen previously, $\lambda^2$ increases with $\d$ and the string tension increases with the increasing effective coupling $\lambda^2$. This dependence of the string tension on the effective coupling strengthens the QCD-like nature of the string tension calculated herein.

In the BPS limit, since $u_2\to 0$, the gravity background \eqref{dcgeometry} reduces to the decoupled geometry of the BPS D$2$ brane and the dual gauge theory becomes $\mathcal{N}=8$ super-Yang-Mills theory in $2+1$ dimensions. Since the super-YM theory is not a confined theory, the string tension $\sigma$ does not exist there. This can be shown here too. In our case, for a moment, assume $\sigma$ exists. Now, if we take $u_2\to 0$, to get a finite turning point in \eqref{tpoint} we need to take $\Sigma\to 1$, i.e., $\g\to 0$ or $\d\to\frac{8}{5}$. Now as $\d\to\frac{8}{5}$, the string tension in \eqref{tension} takes the form $\frac{\sigma}{g_\text{YM}^4N_c^2}=\frac{9\pi^3}{2}\left(\frac{u_2}{L}\right)^\frac{5}{2}$, which goes to zero as $u_2\to 0$. So in the BPS limit the theory becomes deconfined like the $(2+1)$ dimensional $\mathcal{N}=8$ super-Yang-Mills theory. 

The analytic expression \eqref{tension1} shows that the flux tube tension increases with the increasing value of $\lambda$ and $g_\text{YM}^2$ at large $N_c$. So the binding energy of the \qq pair bounded by the flux-tube also increases with these couplings, which is consistent with the expected gauge theory. The linear nature of the potential in \eqref{potential} indicates that the \qq binding energy increases with the increase of their separation. This is the confinement property of the QCD. The existence of the confinement indicates the theory to be in the self-coupling dominated non-perturbative regime. \\

\section{Mass Spectrum of $0^{-+}$}

In this self-coupling dominated regime, the confinement nature of the theory brings the gluons closer and makes the bound states -- glueball. Though there is no complete theoretical model of the non-perturbative strongly coupled gauge theory, an empirical model of the glueball in QCD$4$ has been constructed in the Bag model \cite{Chodos:1974je,Chodos:1974pn}. Although, the glueball is not observed experimentally till date, the mass spectrum of the glueballs has been calculated theoretically \cite{Constable:1999ch,Csaki:2006ji,deMelloKoch:1998vqw} and in lattice QCD \cite{Teper:1998te,Morningstar:1999rf,Chen:2005mg}. Those discrete spectra have also been found in the holographic QCD approaches. The same study leads us to the glueballs in the three dimensional QCD -- both in lattice \cite{Teper:1998te,Teper:1997tq,Philipsen:1996pg} and holographic approaches \cite{Aharony:1999ti,Nair:2002yg,Hong:2010sb}. Here we study the pseudoscalar glueball spectrum in holographic QCD in three dimensions using the non-susy solutions of the type-II supergravity. The pseudoscalar glueball mass is associated with the axion field of the bulk. 

In our bulk theory \eqref{geometry} the axion field is zero. But one can consider the fluctuation of this axion to be non-zero. As the axion field couples minimally with the background, the fluctuation does not change the background metric. Now considering the action for the minimally coupled scalar, the linearised field equation for axion fluctuation $\chi$ can be written as,
\be\label{axioneq}
\frac{1}{\sqrt{-G}}\partial_\mu\left(\sqrt{-G}G^{\mu\nu}\partial_\nu\right)\chi=0
\ee 
where $G$ is the determinant of the background metric $G_{\mu\nu}$\eqref{dcgeometry} and $\mu,\,\nu$ run over all of the ten coordinates of the decoupled geometry. Now for simplicity we demand the fluctuation field $\chi$ to be symmetric on the six dimensional transverse sphere of \eqref{dcgeometry}, polarised along the brane worldvolume, and also a  function of $u$. So we take $\chi=f(u)e^{ik_ax^a}$, where $a=0,\,1,\,2$ and $k_ak^a=-M^2$. $M$ denotes the mass of the pseudoscalar glueball $0^{-+}$. Along with these assumptions using the geometry \eqref{dcgeometry} we get from \eqref{axioneq},
\bea 
&& \partial_u^2f+\left[\frac{6}{u}+\half \frac{\partial_uF}{F}+\left(1+\d-\frac{\b}{4}\right)\frac{\partial_uG}{G}\right]\partial_uf\nn
&& \quad\quad\quad\quad +\frac{LM^2}{\g u_2^5}FG^{-\frac{4}{5}-\frac{\b}{2}}f=0
\eea
As the harmonic functions change very rapidly near the singularity, we take the coordinate transformation $u=u_2e^y$  to properly analyse the whole range of $u$. $y$ is the new dimensionless radial coordinate. The radial coordinate range is now zoomed into $-\infty<y<\infty$ from $0<u<\infty$. Using this new coordinate the above equation can be written as follows.
\bea 
&& \partial_y^2f+\left[5+\half \frac{\partial_yF}{F}+\left(1+\d-\frac{\b}{4}\right)\frac{\partial_yG}{G}\right]\partial_yf\nn
&& \quad\quad\quad\quad +\frac{LM^2}{\g u_2^3}e^{2y}FG^{-\frac{4}{5}-\frac{\b}{2}}f=0
\eea
Substituting $f(y)=e^{-\frac{5}{2}y}F^{-\frac{1}{4}}G^{-\half-\frac{\d}{2}+\frac{\b}{8}}\kappa(y)$ yields the standard form of the differential equation to use the WKB method.
\be 
\partial_y^2\kappa(y)-V(y)\kappa(y)=0
\ee 
This is a second order Schr\"odinger-like wave equation. The potential function is given as,
\bea 
V(y) & = & \frac{25}{4}+\frac{25}{4}\frac{\left(\frac{15}{16}\d-\frac{1}{4}\g+1\right)\left(\frac{15}{16}\d-\frac{1}{4}\g-1\right)}{(e^{5y}+1)^2}\nn
&& +\frac{25}{4}\frac{\g\left(\frac{15}{16}\d+\frac{3}{4}\g\right)}{(e^{5y}+1)^2\left(1-(1+e^{-5y})^{-\g}\right)}\nn
&& -\frac{75}{16}\frac{\g^2}{(e^{5y}+1)^2\left(1-(1+e^{-5y})^{-\g}\right)^2}\nn 
&& -\frac{m^2}{\g}e^{2y}\frac{(1+e^{-5y})^\g-1}{\left(1+e^{-5y}\right)^{\frac{4}{5}+\frac{\g}{2}+\frac{\d}{8}}}
\eea 
where the mass $M$ is made dimensionless quantity $m$ by using $m^2=M^2L/u_2^3$. To analyse the nature of the potential function we consider the asymptotic expansions. The approximated expansion of $V(y)$ at $y\to\infty$ is
\be 
V(y)\approx \frac{25}{4}-m^2e^{-3y}
\ee 
So the potential function gradually reaches to the constant numerical value $25/4$ at positive infinity and the cross-over point in this direction is at $y_+=\frac{2}{3}\ln\left(\frac{2}{5}m\right)$. At the negative asymptote, $y\to-\infty$, the potential function can be approximated as,
\be
V(y)\approx\frac{25}{4}\left(\frac{15}{16}\d+\frac{1}{4}\g\right)^2-\frac{m^2}{\g}e^{(6-\frac{5}{2}\g+\frac{5}{8}\d)y}
\ee
Here also $V(y)$ merges with a positive constant $\frac{25}{4}\left(\frac{15}{16}\d+\frac{1}{4}\g\right)^2$ with a cross-over at $y_-=\frac{16}{48-20\g+5\d}\ln\left(\frac{5}{2}\frac{\sqrt{\g}}{m}\left(\frac{15}{16}\d+\frac{1}{4}\g\right)\right)$. So the potential is almost flat and positive in whole range except $y_-\leq y\leq y_+$. Thus it forms a small potential well with boundaries at $y_-$ and $y_+$. The depth of the well is regulated by $\d$ and the mass $m$. Thus using WKB approximation we can estimate $m$ for given $\d$. The WKB approximation equation is
\be\label{wkb}
\int_{y_-}^{y_+}dy\sqrt{-V(y)}=(n+\half)\pi
\ee
where $n=0,\,1,\,\cdots$ denote various energy states. The integrand in \eqref{wkb} is function of $\d$ and $m$. So, solving this equation, $m$ can be evaluated for the given $\d$ and $n$. This approximation is valid if the depth of the well is small enough. As the depth of the well is directly proportional to the mass $m$, (which can be shown easily by plotting $V$ for different values of $m$) the WKB approximation is a good approximation for the low energy states. Here we calculate first few energy states of the spectrum. Using numerical tools we solve \eqref{wkb} and find the values of $M$ in the unit of $\sqrt{u_2^3/L}$. Then we numerically calculate the ratio $M/\sqrt{\sigma}$ and compare them with some previous results.   

For the scalar glueball $0^{++}$, the Schr\"odinger-like wave equation can be found easily from the linearized equation of the dilaton field in the Einstein frame. In this case, the potential well is not bounded at the negative $y$ regime, i.e., $y_-\to -\infty$. Now as the background geometry \eqref{dcgeometry} has singularity at $y=-\infty$, we need to put a cut-off near the singularity to evaluate the spectrum using WKB approximation \eqref{wkb}. Although, the spectrum is then found to depend on the cut-off. Again the gravity theory near the singularity has the large value of the effective coupling $\lambda^2$ and therefore, the theory has to be uplifted to the eleven dimensional theory. So, to find the scalar glueball spectrum in this present background the WKB approximation method does not work.

Since the analytic turning points $y_+$ and $y_-$ come from the asymptotic expansions ignoring higher order terms, the actual turning points for the whole potential $V(y)$ slightly differ from those values. So in numerical method, first we find those exact turning points numerically. Then using those values, the equation \eqref{wkb} is numerically solved for $m$. 
\begin{table}[h]
    \centering
    \caption{Here we calculate the mass $M_{0^{-+}}$ and $\sqrt{\sigma}$ in unit of $\sqrt{u_2^3/L}$ and $10^{-3}L$ respectively, for various $d$, where $M_{0^{-+}},\,M_{0^{-+}}^*$ and $M_{0^{-+}}^{**}$ denote the mass of the ground state, first excited state and second excited state respectively.}
    \begin{tabular}{|c|c|c|c|c|}
         \hline
         $\d$ & $M_{0^{-+}}$ & $M_{0^{-+}}^*$ & $M_{0^{-+}}^{**}$ & $\sqrt{\sigma}$\footnote{These values have been found by dividing the values in the Table \ref{tab:tension} by $3\pi^2$.}\\
         \hline
         $ 0.98 $ & $ 6.15797 $ & $ 9.70152 $ & $ 12.9685 $ & $ 6.07865 $\\
         $ 1.08 $ & $ 6.25563 $ & $ 9.86602 $ & $ 13.2028 $ & $ 6.45687 $\\
         $ 1.18 $ & $ 6.35031 $ & $ 10.0252 $ & $ 13.4294 $ & $ 6.70183 $\\
         $ 1.28 $ & $ 6.44245 $ & $10.1799 $ & $ 13.6493 $ & $ 6.90446 $\\
         $ 1.38 $ & $ 6.53266 $ & $10.3310 $ & $ 13.8638 $ & $ 7.08193 $\\
         $ 1.48 $ & $ 6.62202 $ & $10.4803 $ & $ 14.0746 $ & $ 7.24190 $\\
         $ 1.58 $ & $ 6.71494 $ & $10.6346 $ & $ 14.2905 $ & $ 7.38863 $\\
         \hline\hline
    \end{tabular}
    \label{tab:mass}
\end{table}
In the Table \ref{tab:mass}, we have listed the dimensionless scaled mass $m$ for different $\d$, where the actual mass $M_{0^{-+}}$ is $m\sqrt{u_2^3/L}$. The mass values have been found to increase with the increasing values of $\d$. We have already seen the effective coupling $\lambda$ as an increasing function of $\d$. So the increasing of $\d$ causes increase in the self-coupling of the gluon fields. Due to this rising of self-coupling, the glueballs become more massive in the sea of gluon. Therefore in this aspect, the calculated mass value shows the consistent nature. The ratio $M_{0^{-+}}^*/M_{0^{-+}}$ are $1.575$ and $1.583$ for $\d=0.98$ and $\d=1.58$ respectively. We can also evaluate the ratio $M_{0^{-+}}/\sqrt{\sigma}$. To do that, we first write $M$ in the unit of $L$ same as $\sqrt{\sigma}$ using $u_2=L/(3\pi^2)$.  From the Table \ref{tab:mass} these ratios are roughly equal to $6,\,9$ and $12$ for the ground state ($n=0$), first excited state ($n=1$) and second excited state ($n=2$) respectively. Thus, using an empirical relation  for $u_2=g_\text{YM}^2N_c$, we can relate the pseudoscalar glueball mass and the QCD string tension as follows.
\be\label{mt_ratio}
M_{0^{-+}}\approx\frac{10^3}{6\sqrt{3}\pi^3}\left(n+2\right)\sqrt{\sigma}\approx 3\left(n+2\right)\sqrt{\sigma}
\ee
where $n=0,\,1\,\&\,2$. In \eqref{mt_ratio}, the first `approximation' sign is used as we are writing this relation for the few states not the whole spectrum and the second `approximation' sign is used simply because of the numerical approximation. Therefore the mass-to-tension ratio is almost constant for a particular energy state, which is one of the similarities of QCD3 with the four dimensional QCD. From this empirical relation the ratio of the mass of the first excited state  to that of the ground state mass is $1.5$. To date we have not found the enough data for the full pseudoscalar spectrum. But the complete spectrum is available in four dimensions. So now we first take a comparative look on those results. In QCD4 \cite{Morningstar:1999rf}, $M_{0^{++}}^*/M_{0^{++}}=1.54$ and $M_{0^{-+}}/M_{0^{++}}=1.50$. On the other hand, in QCD$3$ these two ratios are respectively $1.5061$ and $2.148$ in lattice calculation \cite{Teper:1998te}. However in Holographic approach \cite{Csaki:1998qr,Aharony:1999ti}, $M_{0^{++}}^*/M_{0^{++}}=1.72$ in QCD$3$. So it seems that the glueballs are comparatively more massive in $2+1$ dimensions. Therefore we can expect the ratio $M_{0^{-+}}^*/M_{0^{-+}}$ which is found to be $1.40$ in lattice calculation \cite{Morningstar:1999rf} and $1.46$ in holography \cite{Csaki:1998qr} for QCD$4$ acquires a higher value in the $2+1$ dimensional theory. Here we get the ratio in the range $1.575$-$1.583$. Due to the behavioural consistencies of this mass spectrum with the QCD, we can expect that these masses are also consistent with the pure Yang-Mills theory in $2+1$ dimensions.\\

\section{Conclusion}

To conclude, in this article, we have obtained the string tension and the pseudoscalar glueball spectrum in $2+1$ dimensional strongly coupled gauge theory from the non-susy D$2$ brane of type-IIA supergravity. The gravity background has two free parameters -- the dimensionless $\d$ and the dimension-full $u_2$. $u_2$ controls the nature of the gauge theory and $\d$ measures the zero temperature gluon-condensate in that gauge theory. Here we have taken $u_2\sim g_\text{YM}^2N_c$, which ensures the corresponding gauge theory is strongly coupled $2+1$-dimensional Yang-Mills theory, i.e., the $2+1$-dimensional QCD. Here we would like to mention that in case of the non-susy brane the worldvolume theory may contain some open string tachyonic field \cite{Sen:1999mg,Sen:2004nf}. But, as we have observed that the decoupled gravity background \eqref{dcgeometry}, asymptotically reduce to the near horizon geometry of the BPS D$2$ brane, i.e., this near horizon geometry is some deformation of the near horizon geometry of the BPS D$2$ brane, so our corresponding gauge theory is some perturbation of the worldvolume theory of the BPS D$2$ brane \cite{Lu:2004ms,Nayek:2015tta}. Therefore the gauge theory considered here is a stable $2+1$ dimensional Yang-Mills theory without tachyon\footnote{In the background \eqref{geometry}, when $\theta=0$, the charge of the non-susy D$2$ brane becomes zero. Then the worldvolume theory becomes fully tachyonic. In general for finite $\theta$ the worldvolume theory is the Yang-Mills theory in presence of tachyonic field. It can be shown that for large $\theta$ and $\gamma=1$ the background reduces to a deformation of the near horizon geometry of BPS D$2$ brane which is tachyon-free. Again the background \eqref{geometry} becomes BPS D$2$ brane in the BPS limit. So in this case, the decoupling limit \eqref{dcscale} makes the charge parameter $\theta$ large, which ensures the absence of the tachyon in the worldvolume theory.}. In the Table \ref{tab:tension}, the string tension $\sigma$ is found to depend on the parameter $\d$. This parametric dependence has been shown pictorially in Figure \ref{sigma vs delta}. It can be found that for a particular value of $\d$, the string tension matches exactly with the results of the pure Yang-Mills theory in $2+1$ dimension. In this article we have numerically calculated the mass spectrum of the pseudoscalar glueball using WKB approximation. The spectrum given in the Table \ref{tab:mass} is found to have consistent nature with QCD. The masses increase with the increasing effective coupling of the theory. However the coupling dependence of the spectra has not been shown explicitly. The ratio of the mass to the square root of the string tension has been found to be constant which is analogous to the four dimensional QCD. 

Here in this study the non-AdS form of the decoupled geometry \eqref{dcgeometry} of the non-susy D$2$ brane indicates that the holographic gauge theory is the non-supersymmetric Yang-Mills theory. The effective gauge coupling $\lambda^2$ is monotonically decreasing function of the energy scale $u$ which is shown in Figure \ref{lambda vs u}. We have also seen the string tension and the pseudoscalar guleball spectrum show the consistent behaviours. Also the existence of the parameter $\d$ ensures the gluon condensate in dual gauge theory. All of these similarities indicate that the holographic dual of the non-susy D$2$ brane in the considered parametric range is the $2+1$-dimensional QCD-like theory. But the non-AdS geometrical structure of the gravity background has made it hard to find the complete picture of this QCD-like gauge theory.

Due to the complicated mathematical structure of the background, we have evaluated the mass spectrum numerically using WKB approximation. It will be interesting to find the mass spectrum analytically. In this way, we will be able to find the analytic expressions of the masses showing explicit dependencies on various parameters. In analytic method the singularity will no longer be an issue in case of the scalar glueballs. Therefore, as an extension of this we can find the analytic expression of the scalar and also the higher spin glueballs in this background. The thermal evolution of the string tension and the glueball spectrum in the $2+1$ dimensional gauge theory is also an interesting topic to study in non-susy brane background of type-II supergravity.      

\vspace{0.5cm}

{\em Acknowledgement} -- We want to acknowledge Shibaji Roy and Kamal L Panigrahi for some important discussions to execute this problem. KN wants to acknowledge the Department of Physics, IIT Kharagpur for the funding.

\end{document}